\documentclass[12pt]{iopart}
\usepackage{bm}
\usepackage{graphicx}
\usepackage{color}

\begin{document}

\title{Site dilution and charge disorder effect on physical properties in SrRu$_{1-x}$Ga$_x$O$_3$}

\author{Renu Gupta, Imtiaz Noor Bhatti and A K Pramanik}
\address{School of Physical Sciences, Jawaharlal Nehru University, New Delhi - 110067, India}

\eads{\mailto{akpramanik@mail.jnu.ac.in}}

\begin{abstract}
Here, we report an evolution of structural, magnetic and transport behavior in doped SrRu$_{1-x}$Ga$_x$O$_3$ ($x$ $\le$ 0.2). The nonmagnetic dopant Ga$^{3+}$ (3$d^{10}$) not only acts for magnetic site dilution in SrRuO$_3$ but also it modifies the Ru charge state and electronic density. Our studies show that Ga$^{3+}$ substitution does not affect the original orthorhombic-\textit{Pbnm} structure of SrRuO$_3$ which is due to its matching ionic radii with Ru$^{4+}$. However, Ga$^{3+}$ has a substantial effect on the magnetic behavior of SrRuO$_3$ where it decreases both magnetic moment as well as magnetic transition temperature $T_c$. Further, this dilution induces Griffiths phase behavior across $T_c$ and cluster-glass behavior at low temperature with higher concentration of doping. The magnetic critical exponent $\beta$ increases with $x$ due to this site dilution effect. The Ga$^{3+}$ induces an insulating state in SrRuO$_3$ with $x$ $>$ 0.05. The charge transport in paramagnetic as well as in insulating state of samples can be well described with Mott's modified variable-range-hopping model. The metallic charge transport just below $T_c$ in SrRuO$_3$ obeys Fermi liquid behavior which, however breaks down at low temperature. We further find a correlation between field dependent magnetoresistance and magnetization through power-law behavior over the series.
\end{abstract}


\submitto{\JPCM}

\maketitle
\section {Introduction}
The 4$d$ based oxide SrRuO$_3$ is commonly believed to be an itinerant ferromagnet (FM) with transition temperature $T_c$ $\sim$ 163 K. \cite{allen,cao,fuchs,mazin,kim-li,cheng,gupta1, gupta2} The itinerant character of magnetism is very evident in the fact that the observed magnetic moment ($\sim$ 1.4 $\mu_B$/f.u), even measured in high enough magnetic field around 30 T, comes lower than its expected spin-only value, 2 $\mu_B$/f.u. \cite{cao} The nature of magnetism in SrRuO$_3$ closely follows the mean-field type, \cite{fuchs,cheng,kim-mf, gupta1, gupta2} however, recent calculations as well as experimental data indicate that this material probably hosts both itinerant and localized type of magnetism. \cite{kim-li,gupta1} The electrical transport behavior in SrRuO$_3$ is equally interesting. The electrical resistivity ($\rho$) increases continuously even at high temperature till $\sim$ 1000 K. However $\rho (T)$ exhibits a distinct slope change across $T_c$ that establishes its close correlation with the magnetic behavior. \cite{allen} The SrRuO$_3$ is considered to be a `bad metal' as $\rho (T)$ at high temperature crosses the Ioffe-Regel limit that is considered for (good) metallic conductivity. \cite{allen, gurvitch, klein1, emery} A Fermi-liquid (FL) type charge conduction at low temperature and its breakdown at high temperature below $T_c$ have been shown in SrRuO$_3$ in several studies. \cite{klein1, wang, klein-b, tyler, hussey, bruin, schneider} The x-ray photoemission spectroscopy (XPS) studies have further shown a contradicting results favoring both significant as well as weak electron correlation strength ($U$) in SrRuO$_3$.\cite{kalo-xps, fujioka} In spite of large volume of studies, the interrelation between magnetism and transport behavior in SrRuO$_3$ remains poorly understood. The high temperature metallic behavior of SrRuO$_3$ has enabled this material to be used as electrode in fuel cell an other devices. Further, due to its high spin polarization factor, SrRuO$_3$ has potential application in spintronics. Recent observation of magnetic skyrmion phase in ultrathin films of SrRuO$_3$ has opened another avenue for its future technological applications.\cite{wang1,ohuchi} Based on its immense importance from both basic and applied sciences, this material continues to attract attention in condensed matter community.

To understand the complex properties of SrRuO$_3$, the route of chemical substitution at both Sr- and Ru-site has been adopted in various studies. The isoelectronic substitution in Sr$_{1-y}$Ca$_y$RuO$_3$ causes a suppression of FM state, and around $y$ = 0.7 the system shows a quantum phase transition (QPT). \cite{cao,fuchs} Regarding Ru-site doping, two types of elements are of obvious choice; magnetic (Fe$^{4+}$, Rh$^{4+}$, Ir$^{4+}$, Cr$^{4+}$, etc) and nonmagnetic (Ti$^{4+}$, Pb$^{4+}$, Zn$^{2+}$, Mg$^{2+}$). \cite{yamaura,dabrowski,kasinathan,qasim,biswas,bianchi,jang,kim-mit,kim-ph,lin-nd,cao-n,pi,crandles,maiti} While former type participates in existing magnetic interactions, the later causes a dilution in magnetic lattice. In the category of nonmagnetic doping, substitution of Ti$^{4+}$ has been well studied. \cite{bianchi,jang,kim-mit,kim-ph,lin-nd,maiti} A recent study by us has shown that Ti$^{4+}$ substitution (till 70\%) weakens the magnetic moment and the strength of its interaction, though the FM $T_c$ remains largely unaffected that has been explained with opposite changes of $U$ and density of states $N(\epsilon_F$) under the model of itinerant ferromagnetism.\cite{gupta1}

In present study, we have used another nonmagnetic substitution Ga$^{3+}$ (3$d^{10}$) for Ru$^{4+}$ in SrRu$_{1-x}$Ga$_x$O$_3$ ($x$ $\leq$ 0.2). The Ga$^{3+}$ has matching ionic radii (0.62 \AA) with Ru$^{4+}$ (0.62 \AA) similar to Ti$^{4+}$ (0.605 \AA), which not only minimizes the effect of structural modification but also ensures it will substitute the Ru$^{4+}$. The introduction of Ga$^{3+}$ in magnetically and electrically active Ru-O-Ru channel will act for site dilution similar to another nonmagnetic element Ti$^{4+}$ (3$d^0$). However, Ga$^{3+}$ will have added effects where it will convert an equivalent amount of Ru$^{4+}$ to Ru$^{5+}$ due to charge balance mechanism. Hence, there would be a mixture of Ru$^{4+}$ and Ru$^{5+}$ ions which will influence the magnetic and electronic properties accordingly. Further, the effective electron density will be modified in an opposite way in case of Ti$^{4+}$ and Ga$^{3+}$ substitution, therefore, the Stoner criteria, $UN(\epsilon_F)$ $>$ 1, for itinerant FM will be affacted. While there are some reports of Ti doped SrRuO$_3$ studies,\cite{gupta1} to the best of our knowledge, this is the first study of Ga substitution in SrRuO$_3$. The results will help to widen our knowledge about the effect of nonmagnetic substitution in SrRuO$_3$.
 
Our studies show Ga substitution does not induce any structural phase transition but the structural parameters evolve with doping. Both $T_c$ as well as magnetic moment decreases with $x$. As a consequence of magnetic site dilution, system develops Griffiths phase and cluster glass like behavior at higher doping concentration. A metal to insulation transition is induced for $x$ $>$ 0.05. The charge conduction in insulating part of samples follow a modified Mott's variable-range-hopping model while in metallic phase it follows FL type behavior. We further observe a correlation between magnetoresistance and magnetization in samples. 

\begin{figure}
	\centering
		\includegraphics[width=8cm]{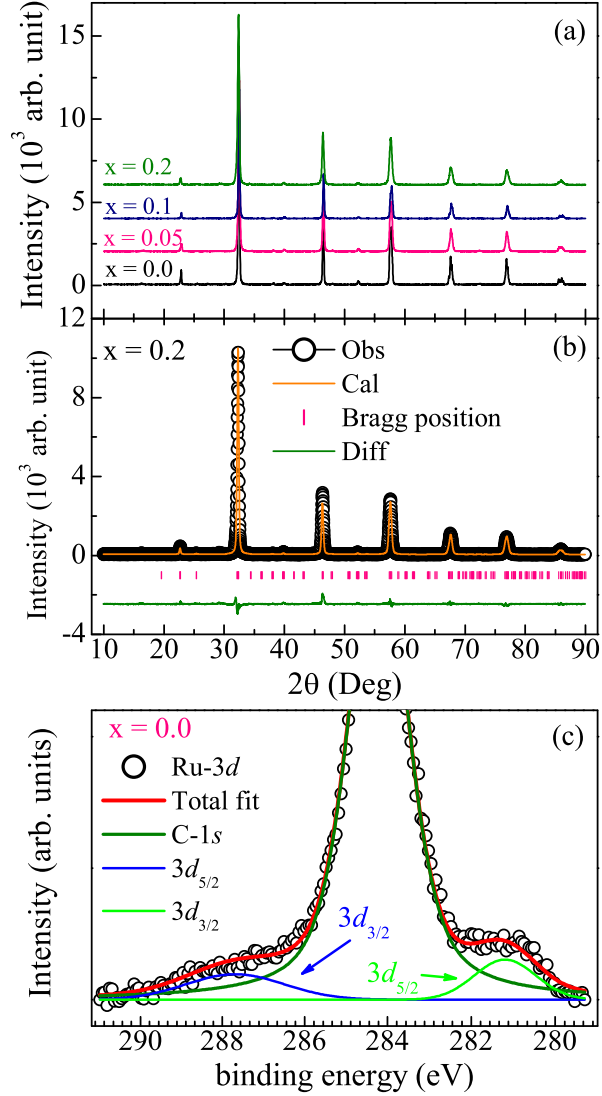}
	\caption{(a) The room temperature powder XRD patterns of SrRu$_{1-x}$Ga$_x$O$_3$ series with $x$ = 0.0, 0.05, 0.1 and 0.2 compositions. (b) The XRD pattern are shown along with Rietveld refinement representation for SrRu$_{0.8}$Ga$_{0.2}$O$_3$ which has orthorhombic-\textsl{Pbnm} structure at room temperature. (c) shows Ru-3$d$ XPS data for SrRuO$_3$ ($x$ = 0.0 along with fittings.)}
	\label{fig:Fig1}
\end{figure} 

\section {Experimental Details}
Polycrystalline series of samples SrRu$_{1-x}$Ga$_x$O$_3$ with nominal compositions $x$ = 0.0, 0.05, 0.1 and 0.2 have been synthesized using solid state reaction method. High purity ($>$ 99.99\% from Sigma Aldrich) ingredient materials, such as SrCO$_3$, RuO$_2$ and Ga$_2$O$_3$ are mixed well in stoichiometric ratio and then heated twice at 1000 $^o$C for 24 hours with an intermediate grinding. Then, powders are pressed into pellets and heated at 1100 $^o$C for 36 hours. Details of sample preparation and analysis are given elsewhere. \cite{gupta1} XRD data are collected using Rigaku MiniFlex diffractometer with Cu$K_\alpha$ radiation in the 2$\theta$ range from 10 to 90$^o$ at room temperature. The recorded XRD patterns are analyzed following Rietveld refinement program by FullProf software.\cite{carvajal} The x-ray photoemission spectroscopy (XPS) measurements have been done at the base pressure of $\sim$ 10$^{-10}$ mbar using a commercial electron energy analyzer (Omicron nanotechnology) attached with non-monochromatic Al-$K_{\alpha}$ x-ray source ($h\nu$ = 1486.6 eV). For the XPS measurement, sample is used in pellet form where the ion beam sputtering has been used before measurement for cleaning the sample surface. XPS data have been analyzed using XPS peakfit 4.1 software. Temperature ($T$) and magnetic field ($H$) dependent magnetization ($M$) data are collected using a superconducting quantum interference device (SQUID) magnetometer by Quantum Design. The electrical resistivity and magnetoresistance of present series is measured by a home-made instrument using the four-probe method. The ac magnetic susceptibility have been measured using a home-made setup in temperature range down to liquid nitrogen (LN$_2$) temperature where the temperature and induced signals are collected using a temperature controller (LakeShore Cryotronics) and a lock-in amplifier (Stanford Research Systems).

\begin{table}
\caption{\label{tab:table I} The structural parameters of SrRu$_{1-x}$Ga$_x$O$_3$ series are shown in terms of lattice parameters $a$, $b$, $c$, unit cell volume ($V$), basal bond-length (Ru-O1), apical bond-length (Ru-O2), basal bond angle (Ru-O1-Ru) and apical bond angle (Ru-O2-Ru)}.
\begin{indented}
\item[]\begin{tabular}{cccccc}
\hline
Ga(x) &0.0 &0.05 &0.1 &0.2\\
\hline
$a$ (\AA) &5.5733(2) &5.5730(4) &5.5678(2) &5.5662(4)\\
\hline
$b$ (\AA) &5.5336(2) &5.5332(3) &5.5311(2) &5.5307(4)\\
\hline
$c$ (\AA) &7.8475(3) &7.8473(3) &7.8377(3) &7.8293(7)\\
\hline
$V$ ({\AA}$^3$) &242.02(1) &241.98(4) &241.37(2) &241.03(3)\\
\hline
Ru-O1 (\AA) &1.9984(1) &1.9936(1) &1.9928(1) &1.9789(1)\\
\hline
Ru-O2 (\AA) &1.9665(1) &1.9657(1) &1.9613(1) &1.9587(2)\\
\hline
Ru-O1-Ru &158.482(1) &159.489(2) &159.772(2) &164.865(3)\\
\hline
Ru-O2-Ru &172.148(1) &173.152(1) &174.945(1) &175.701(2)\\
\hline
$R_{wp}$ &14.1 &14.0 &15.1 &12.9\\
\hline
$R_p$ &12.1 &10.9 &11.1 &10.4\\
\hline
\end{tabular}
\end{indented}
\end{table}

\section{Results and Discussions}
\subsection{Structural and chemical analysis}
The room temperature XRD patterns of present series are shown in Fig. 1a. As seen in the figure, XRD pattern do not show any modification with Ga composition in terms of peak position or any additional peak. We have previously shown that parent SrRuO$_3$ crystallizes in GdFeO$_3$-type orthorhombic \textit{Pbnm} structure in agreement with reported studies. \cite{cao} We find that the original orthorhombic-\textit{Pbnm} structure is continued for present whole series. Fig. 1b shows representative Rietveld analysis of XRD data for doped $x$ = 0.2 sample which is the last member of the present series. The Rietveld refinement shows a reasonably good fitting of XRD data for SrRu$_{0.8}$Ga$_{0.2}$O$_3$ with original orthorhombic-\textit{Pbnm} structure. Here, it can be mentioned that we obtain fitting indicator such as, $\chi^2$ value about 1.74, 1.97, 1.41 and 1.74 for $x$ = 0.0, 0.05, 0.1 and 0.2, respectively of present series which suggests a reasonably good fitting. Our Rietveld fitting is further evaluated with \textit{Goodness of fitting} parameter ($R_{wp}$/$R_p$) parameter which yields values within 1.3 for the whole series, implying a good fitting (Table I). Given that Ga$^{3+}$ and Ru$^{4+}$ has matching ionic radii (0.62 \AA), therefore it is least expected for any major structural modification in the present series.

Table 1 shows the evolution of structural parameters with Ga substitution which are obtained from Rietveld analysis. The lattice parameters $a$, $b$ and $c$ and volume $V$ show that the values decrease with Ga doping concentration. This evolution in lattice parameters can be understood from ionic distribution. The substitution of Ga$^{3+}$ equivalently generates Ru$^{5+}$, which has comparatively lower ionic radii  (0.565 \AA) than Ru$^{4+}$. Therefore, decreasing lattice parameters affect the unit cell volume, $V$. We have calculated the percentage change in structural parameters of SrRu$_{1-x}$Ga$_{x}$O$_3$ series such as, $\Delta a$, $\Delta b$, $\Delta c$ and $\Delta V$ with values around 0.007, 0.003, 0.018 and 0.99\%, respectively. The changes in respective values are very insignificant which follows our expectation due to matching ionic radii. We also observe the change in RuO$_6$ octahedral parameters. The SrRuO$_3$ adopts a GdFeO$_3$-type crystal structure where RuO$_6$ octahedra demonstrates distortion both along ab-plane as well as along $c$-direction which causes the basal and apical Ru-O-Ru bond angle to deviate from the ideal condition. Table 1 shows that both the basal Ru-O1 and apical Ru-O2 bond-length decreases in agreement with changes in lattice parameters. The basal ($<$Ru-O1-Ru$>$) as well as apical ($<$Ru-O2-Ru$>$) bond-angle for SrRuO$_3$ in Table 1 agree with the reported values and implies RuO$_6$ octahedra are both rotated and tilted along $c$-axis. With Ga substitution, both the bond-angle increases toward ideal value of 180$^o$, suggesting octahedral distortion is relieved progressively. 

\begin{figure}
	\centering
		\includegraphics[width=8.5cm]{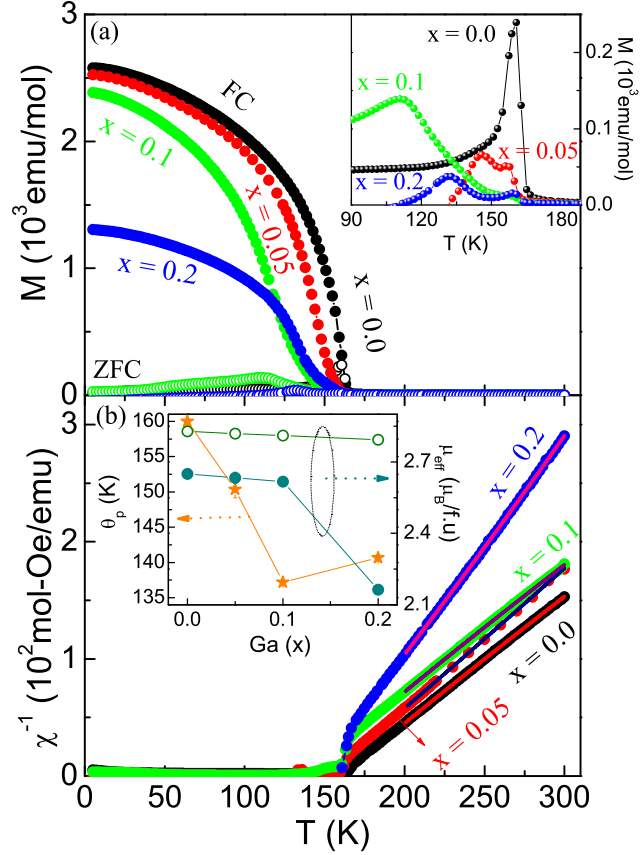}
	\caption{(a) Temperature dependent ZFC and FC dc magnetization measured in 100 Oe field are shown as a function of temperature for SrRu$_{1-x}$Ga$_x$O$_3$ series. The inset shows $M_{ZFC}$ data close to $T_c$. (b) The inverse susceptibility ($\chi^{-1}$ = ${M/H}$) are plotted as a function of temperature for the present series where the solid line are due to fitting with modified Curie-Weiss behavior (Eq. 1). In inset, left axis shows the CW temperature ($\theta_P$) and right axis shows the effective PM moment (filled circle experimental data and empty circle calculated data) with Ga doping for SrRu$_{1-x}$Ga$_x$O$_3$ series.}
	\label{fig:Fig2}
\end{figure} 

Elemental analysis of transition metal charge state has been done with x-ray photoemission spectroscopy (XPS). Fig. 1c shows the Ru-3$d$ core level spectra for parent SrRuO$_3$. Along with centrally located dominant peak at 284.2 eV due to C-1$s$ state, the Ru-3$d$ spectra shows two distinct peaks on both side at binding energies 281.2 and 287.7 eV which correspond to spin-orbit split Ru-3$d_{5/2}$ and Ru-3$d_{3/2}$ electronic states, respectively. The observed C-1$s$ peak is mostly due to adhesive carbon tape used in the experiment. The red line in Fig. 1c is due to overall fitting while the respective fittings of C-1$s$ and Ru-3$d$ states are shown in figure. This fitting confirms the transition metal is in Ru$^{4+}$ charge state which is also expected from its chemical composition.\cite{galal,lin}   

\subsection{DC Magnetization study for SrRu$_{1-x}$Ga$_x$O$_3$ series}
The temperature dependent magnetization data, measured from 5 K to 300 K in 100 Oe following zero field cooling (ZFC) and field cooling (FC) protocol, are shown in Fig. 2a for SrRu$_{1-x}$Ga$_x$O$_3$ series. The SrRuO$_3$ is a well documented itinerant type FM with $T_c$ $\sim$ 163 K. \cite{cao} Inset of Fig. 2a shows a close view of $M_{ZFC}(T)$ for all samples across $T_c$. We observe that $M_{ZFC}$ for parent SrRuO$_3$ shows a sharp peak around 163 K and this is an onset temperature of bifurcation between $M_{ZFC}$ and $M_{FC}$ magnetization. In fact, SrRuO$_3$ exhibits large bifurcation between ZFC and FC magnetization data which is suggestive of large anisotropy in this material. \cite{klein2} In doped samples, while the onset temperature for bifurcation remains almost the same, but the amount of bifurcation reduces substantially. These results are similar to Ti doped samples where recently we have shown that the long-range ordering temperature $T_c$ remains almost unchanged till 70\% of Ti substitution which has been explained from opposite tuning of electron correlation and density of states with Ti doping. \cite{gupta1} The present series, however, differs from the Ti doped one as Ga$^{3+}$ not only act for site dilution but it creates Ru$^{5+}$ which is magnetic and participates in magnetic interaction with existing Ru$^{4+}$ ions. It is evident in inset of Fig. 2a that below $T_c$, the $M_{ZFC}$ in doped materials develops an extra peak which may originate from magnetic clustering effect (discussed later).

To understand the high temperature paramagnetic (PM) state, inverse susceptibility ($\chi^{-1}$ = $(M/H)^{-1}$) as deduced from magnetization data in Fig. 2a, is shown of Fig. 2b. The $\chi^{-1}(T)$ in PM regime shows a linear behavior for all the samples which implies a Curie-Weiss (CW) type behavior. The $\chi^{-1}(T)$, however, shows a sharp downturn close to $T_c$ in higher doped samples which are analyzed in terms of Griffiths phase behavior (discussed later). The $\chi^{-1}(T)$ data have been analyzed with modified CW law as following,

\begin{eqnarray}
	\chi = \chi_0 + \frac{C}{T - \theta_P}
\end{eqnarray}   
          
where $\chi_0$ is the temperature independent susceptibility, $C$ is the CW constant and $\theta_P$ is CW temperature. The fitting of Eq. 1 with $\chi^{-1}(T)$ data have been shown in Fig. 2b down to 200 K indicating reasonably good fitting. The fitting yields $\chi_0$ = 2.56 $\times$ 10$^{-4}$ emu mol$^{-1}$Oe$^{-1}$, $C$ = 0.88 emu K mole$^{-1}$ Oe$^{-1}$ and $\theta_P$ = 160 K for SrRuO$_3$. The positive sign of $\theta_p$ indicates the interaction between Ru ions is of FM type. Using the above value of CW constant, we have calculated the effective PM moment ($\mu_{eff}$) which is about 2.65 $\mu_B$/f.u. for SrRuO$_3$ and closely matches with the reported values \cite{cao, kalo-xps, shepard, miao}. The obtained value is also close to the calculated spin-only value ($g\sqrt{S(S+1)} \mu_B$) 2.83 $\mu_B$/f.u. with $S$ = 1. The inset of Fig. 2b shows the variation of $\theta_P$ (left axis) and $\mu_{eff}$ (right axis) with $x$. The $\theta_P$ initially decreases till $x$ = 0.1 but then shows a slight increase for $x$ = 0.2. While $\theta_P$ remains positive throughout the series but this nonmonotonic behavior may be related to the change in Ru$^{4+}$/Ru$^{5+}$ ratio with $x$. The measured $\mu_{eff}$, on the other hand, decreases linearly till $x$ = 0.1 and then shows a steep decrease. As discussed, doped Ga$^{3+}$ ($S$ = 0) converts Ru$^{4+}$ ($S$ = 1) to Ru$^{5+}$ ($S$ = 3/2). Therefore, the overall $\mu_{eff}$ of the sample can be expressed as, $\mu_{eff}$ = $\sqrt{[(1 - 2x) (\mu_{eff}^{S=1})^2 + x (\mu_{eff}^{S=3/2})^2 + x (\mu_{eff}^{S=0})^2]}$. The calculation shows that overall $\mu_{eff}$ decreases continuously which agrees with our experimental values of $\mu_{eff}$ (inset of Fig. 2b). 

\begin{figure}
	\centering
		\includegraphics[width=8cm]{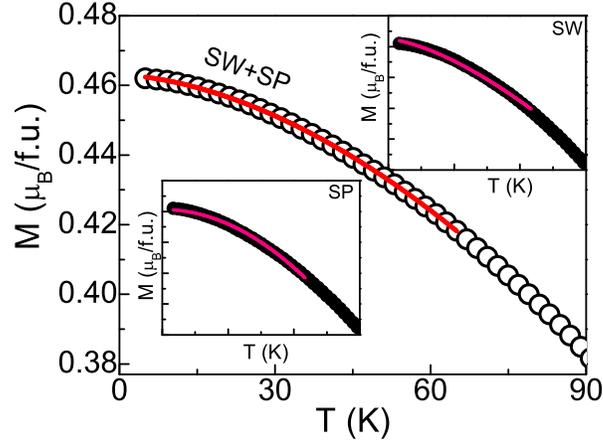}
	\caption{The combined spin-wave and single-particle (Eq. 6) fitting of magnetization data are shown at low temperature for SrRuO$_3$. The upper inset shows the same with only spin-wave fitting (Eq. 2) while lower inset shows with only single-particle fitting (Eq. 5).}
	\label{fig:Fig3}
\end{figure}

\begin{figure}
	\centering
		\includegraphics[width=8.5cm]{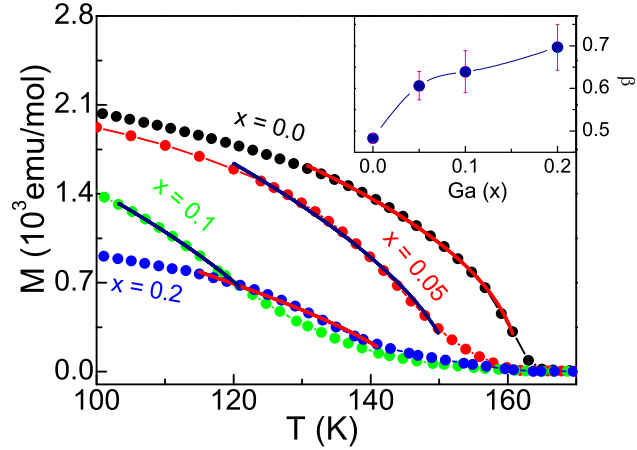}
	\caption{The fitting of $M (T)$ data with Eq. 8 are shown for SrRu$_{1-x}$Ga$_x$O$_3$ series close to $T_c$. The inset shows a variation of exponent $\beta$ with $x$.}
	\label{fig:Fig4}
\end{figure} 

\subsection{Thermal Demagnetization study}
The thermal demagnetization of FM materials at low temperature is of particular interest. Usually, for exchange interaction driven FM systems, the thermal demagnetization is explained with Bloch's spin-wave (SW) theory analysis which is expressed as following (neglecting higher order terms), \cite{kittel}

\begin{eqnarray}
	M (T) = M (0)[1 - BT^{3/2}]
\end{eqnarray}

where $M (0)$ is the zero temperature magnetization and $T^{3/2}$ term comes due to long wavelength. Using the spin-wave coefficient $B$ (Eq. 2), spin-wave stiffness constant $D$ can be calculated as;

\begin{eqnarray}
	D = \frac{k_B}{4\pi}\left[\frac{\zeta(3/2) g\mu_B}{M (0)\rho B}\right]^{2/3}
\end{eqnarray}

where $k_B$, $g$, $\mu_B$ and $\rho$ are the Boltzmann constant, Lande g-factor, Bohr magneton and density of material, respectively and $\zeta$(3/2) = 2.612. Similarly, Stoner single-particle (SP) excitations can be used to understand the low temperature demagnetization of itinerant FM where the origin of magnetism is associated with the splitting of band into spin-up and spin-down sub-bands separated by an exchange energy. The energy difference between two sub-bands is proportional to spontaneous magnetization ($M_s$). The generalized form of Stoner's single-particle excitations is given below, 

\begin{eqnarray}
	M (T) = M (0)\left[AT^k\exp(-\frac{\Delta}{k_BT})\right]
\end{eqnarray}
 
where $A$ is the single-particle excitation coefficient, and $\Delta$ is the energy gap between the top of full sub-band and Fermi level. The itinerant FM are divided into two categories based on $k$ and $\Delta$. For example, $k$ = 2 and $\Delta$ = 0 gives a weak-type itinerant FM where both the sub-bands are partially filled and the Fermi level stays within both sub-bands. On the other hand, a strong-type itinerant FM is realized for $k$ = 3/2 and $\Delta$ $\neq$ 0 where one sub-band is partially filled and another one is fully filled. As the spin-polarization of SrRuO$_3$ is reported less than 100\%, \cite{nadgorny, raychaudhuri} therefore a weak-type itinerant FM is expected in this material with following temperature dependence, 

\begin{eqnarray}
	M (T) = M (0)\left[AT^{2}\right]
\end{eqnarray}
					
\begin{figure}
	\centering
		\includegraphics[width=8cm]{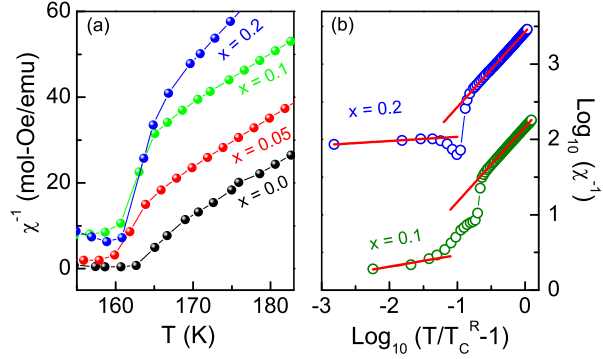}
		\caption{(a) The $\chi^{-1}$ with the variation of temperature are shown for SrRu$_{1-x}$Ga$_x$O$_3$ series across $T_c$. (b) The log$_{10}$ - log$_{10}$ scale plotting of $\chi^{-1}$ data following Eq. 9 are shown for SrRu$_{1-x}$Ga$_x$O$_3$ series with $x$ = 0.1 and 0.2. The straight lines are due to fitting. This plot of $x$ = 0.2 sample has shifted upward by 1 for clarity.}
	\label{fig:Fig5}
\end{figure}   

We have tried to fit the low temperature $M (T)$ data, using both spin-wave (SW) and single-particle (SP) model separately that are shown in insets of Fig. 3. However, the thermal demagnetization data can be better fitted using combined SW and SP model which takes account both the localized as well as an itinerant model of magnetism as given below,

\begin{eqnarray}
	M (T) = M (0)\left[1 - BT^{3/2} - AT^2\right]
\end{eqnarray}

Our recent study has shown that low temperature thermal demagnetization in SrRu$_{1-z}$Ti$_z$O$_3$ can be better explained with combined SW and SP model which is also in line with a recent theoretical calculation that suggests a coexistence of localized and itinerant magnetism in SrRuO$_3$. \cite{gupta1,kim-li} Fig. 3 shows the temperature dependent magnetization data where solid lines are due to the best fitting of $M (T)$ data with Eq. 6 up to $\sim$ 65 K where the fitting of $M (T)$ data is shown with only spin-wave model [Eq. 2, upper inset], single-particle model [Eq. 5, lower inset] and combined model [Eq. 6, main panel] for SrRuO$_3$. As evident in the figure, the Eq. 6 gives better fitting in wide temperature range which is also validated by higher $R^2$ value. The fitting in Fig. 3 (main panel) gives parameters of constants $M (0)$, $B$ and $A$ to be 0.463 $\mu_B$/f.u., 9.0 $\times$ 10$^{-5}$ K$^{-3/2}$ and 1.17 $\times$ 10$^{-5}$ K$^{-2}$, respectively for SrRuO$_3$. The spin-wave stiffness constant $D$ has been calculated using the fitted parameter $B$ using Eq. 6. We obtain $D$ = 264.9 meV $\AA^2$ for SrRuO$_3$. The magnetic exchange coupling constant ($J$) between nearest-neighbor magnetic atoms has also been calculated using following equation \cite{kittel}

\begin{eqnarray}
	J = \frac {k_B} {2S} \left[\frac {0.0587} {BS}\right]^{2/3}
\end{eqnarray}

where $S$ (=1) is the localized atomic spin for SrRuO$_3$. The value of $J$ comes out to be 37.6 $k_B$ K. For the doped materials, we find similar combined model is required to understand the low temperature demagnetization data. According to band magnetism model, \cite{perumal} if the spontaneous moment $M_s$ decreases, then the exchange splitting decreases by an amount, $\Delta$E = $I$$\Delta$$M_s$ /N$\mu_B$, where N is the number of spins per unit volume and $I$ is the Stoner parameter of spin-up and spin-down $d$ sub-bands. In present series, the $M_s$ decreases with Ga doping (inset, Fig. 6b), therefore, it is believed that the Fermi level $E_F$ will be further pushed inside $d_{\downarrow}$ and $d_{\uparrow}$ sub-bands. This suggests both spin-wave, as well as single-particle models would be required to explain the thermal demagnetization effect. The calculated $D$ and $J$ have been shown in Table 2 for present SrRu$_{1-x}$Ga$_x$O$_3$ series. The SW stiffness constant $D$ decrease till $x$ $\sim$ 0.1 and then it increases for $x$ = 0.2. Here, it can be mentioned that increasing $D$ value has been seen in other SrRu$_{1-z}$Ti$_z$O$_3$ series. This opposite evolution of $D$ in both series with nonmagnetic doping is believed to arise due to change of Ru$^{4+}$ ionic state in present series where in Ti doped series the Ru$^{4+}$ charge state does not alter with doping. However, an increase of $D$ in $x$ = 0.2 sample may be due to an interplay between site dilution caused by Ga$^{3+}$ doping and changing ionic state of Ru from Ru$^{4+}$ to Ru$^{5+}$.  

\subsection{Critical exponent behavior}
The nature of magnetic interaction in magnetic materials is usually characterized by a universality class which is associated with different set of critical exponents. \cite{stanley} These universality classes do not depend on microscopic details of materials; rather those are defined by lattice dimension and spin dimension. The temperature dependent magnetization in vicinity of $T_c$ obeys following relation,

\begin{eqnarray}
	M = M_0(T_c - T)^\beta
\end{eqnarray}

where $\beta$ is the critical exponent, and $M_0$ is the magnetization at 0 K. The $M (T)$ data have been fitted with Eq. 8 close to $T_c$ (Fig. 4). In fitting, the $M_0$, $T_c$ and $\beta$ have been kept as free parameters. The obtained values of $M_0$, $T_c$ and $\beta$ for SrRuO$_3$ are 309.57 emu mol$^{-1}$, 161.2(3) K and 0.48(2), respectively, where $\beta$ value implies a mean-field type behavior in SrRuO$_3$. This exponent is consistent with other reported studies. \cite{fuchs,cheng,kim-mf} The values of $\beta$ as a function of Ga doping are shown in inset of Fig. 4. It is clear in the figure that the $\beta$ increases monotonically with Ga, reaching the value $\sim$ 0.696 $\pm$ 0.05 for $x$ = 0.2 in SrRu$_{1-x}$Ga$_x$O$_3$ series. The similar behavior has also been obtained in SrRu$_{1-z}$Ti$_z$O$_3$ series. This increase in $\beta$ with Ga substitution is likely caused by formation of FM clusters across $T_c$, as explained in case of Ti doped in SrRuO$_3$. \cite{gupta1, gupta2}

\begin{table}
\caption{\label{tab:table II} The magnetization parameters ($T_c$, $D$, $J$, $M_r$, $q_c$/$q_s$) and Griffiths phase parameters ($T_c^R$, $T_G$, $\lambda_{PM}$, $\lambda_{GP}$) variation by the Ga substitution of SrRu$_{1-x}$Ga$_x$O$_3$ series.}

\begin{indented}
\item[]\begin{tabular}{cccccc}
\hline
Ga(x) &0.0 &0.05 &0.1 &0.2\\
\hline
$T_c$ (K) &161 &146 &111 &133\\
\hline
$D$ (meV $\AA^2$) &264.9 &206.6 & 179.2 &327\\
\hline
$J$ ($k_B$ K) &37.6 &28.9 &23.1 &28.2\\ 
\hline
$M_r$ ($\mu_B$/f.u) &0.69 &0.61 &0.55 &0.38\\
\hline
$q_c$/$q_s$ &1.46 &1.63 &1.81 &1.82\\
\hline
$T_c^R$ (K) & & &134 &144.7\\
\hline
$T_G$ (K) & & &169 &173\\
\hline
$\lambda_{PM}$ & & &0.004 &0.002\\
\hline
$\lambda_{GP}$ & & &0.846 &0.944\\
\hline
\end{tabular}
\end{indented}
\end{table}

\subsection{Griffiths Phase behavior}
The Griffiths phase (GP) like behavior originates from local disorder present in crystallographic structure and magnetic interactions. \cite{griffiths} The GP behavior is mainly characterized by formation FM clusters at temperatures higher than the $T_c$. The inverse magnetic susceptibility ($\chi^{-1}$) plotted as a function of temperature shows a sudden downturn near $T_c$ in the case of GP. \cite{magen, pramanik-GP, neto} Fig. 5a shows $\chi^{-1}(T)$ within limited temperature regime for SrRu$_{1-x}$Ga$_x$O$_3$ series. The $\chi^{-1}(T)$ data for higher doped samples show a distinct downturn just above $T_c$ which implies a signature of GP like behavior. The temperature where $\chi^{-1}(T)$ data show an onset of downturn is termed as Griffiths temperature ($T_G$) and the temperature range between $T_G$ and $T_c$ is called GP regime where the system neither exhibits ideal FM order nor PM behavior, rather in this regime there exists ferromagnetically ordered finite-size clusters embedded in PM background. Due to the presence of clusters, the magnetization shows nonanalytic behavior, hence the susceptibility diverges which is reflected in a sharp downturn in $\chi^{-1}$. \cite{magen, pramanik-GP} The sudden and sharp downfall of $\chi^{-1}(T)$ above $T_c$ in the GP regime is characterized by following relation: \cite{neto}

\begin{eqnarray}
	\chi^{-1} = (T - T_c^R)^{1-\lambda}
\end{eqnarray}	

where $T_c^R$ is the random critical temperature where the magnetic susceptibility diverges which generally lies between the $T_c$ and $T_G$. 
The choice of proper $T_c^R$ is very important for the correct evaluation of exponent $\lambda$. The exponent $\lambda$ and $T_c^R$ have been determined following the protocol described in Ref. 42. This power law behavior appears a modified form of the CW equation where the finite $\lambda$ arises due to FM cluster formation. The $\lambda$ value lies in the range of 0 to 1. In the PM region the value of $\lambda$ is realized to be 0, hence the Eq. 9 reduces to the standard CW equation. In the GP regime the $\chi^{-1}$ follows the modified CW behavior i.e., Eq. 9 with non-zero value of $\lambda$. Fig. 5b shows semi-log plotting of Eq. 9 both in PM as well as in GP regime for $x$ = 0.1 and 0.2. The straight line fitting in Fig. 5b confirms GP like behavior in present series, as similar to previous Ti doped samples. \cite{gupta1} The obtained values of $T_c^R$, $\lambda_{GP}$ and $\lambda_{PM}$ of doped samples are given in Table 2 while the exponent $\lambda_{PM}$ in PM is close to zero, the $\lambda_{GP}$ shows very high values which indicates a strong behavior of GP in these materials. The validity of GP confirms a formation of short-range FM clusters due to non-magnetic Ga$^{3+}$ substitution and an increase of $\lambda_{GP}$ indicates the sizes of FM clusters decreases with Ga doping.  

\begin{figure}
	\centering
		\includegraphics[width=8.5cm]{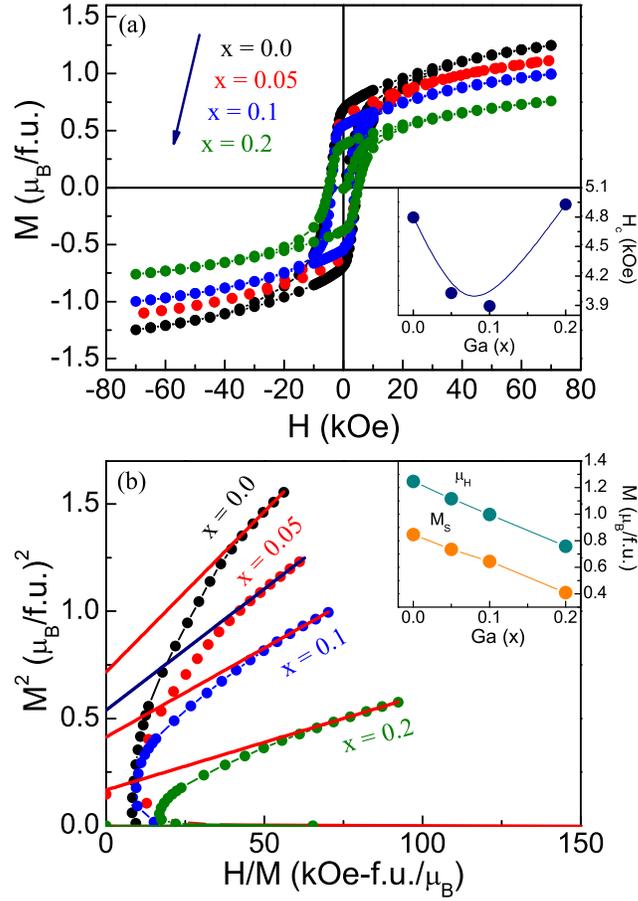}
	\caption{(a) The magnetic field dependent isothermal magnetization at 5 K are shown for SrRu$_{1-x}$Ga$_x$O$_3$ series with $x$ = 0.0, 0.05, 0.1 and 0.2. The inset shows the variation of coercive field ($H_c$) with doping in SrRu$_{1-x}$Ga$_x$O$_3$ series (b) The Arrott plot ($M^2$ vs ${H/M}$) of magnetization $M (H)$ data at 5 K for SrRu$_{1-x}$Ga$_x$O$_3$ series and inset shows the saturation magnetization ($\mu$$_H$) at 7 T and spontaneous magnetization ($M_s$) by Ga doping in SrRu$_{1-x}$Ga$_x$O$_3$ series.}
	\label{fig:Fig6}
\end{figure}
  
\subsection{Isothermal Magnetization study}
Fig. 6a shows the field dependent isothermal magnetization $M (H)$ data collected at 5 K for SrRu$_{1-x}$Ga$_x$O$_3$ series. The data have been collected with a field range of $\pm$ 70 kOe. The parent SrRuO$_3$ shows large hysteresis with coercive field ($H_c$) $\sim$ 4800 Oe which matches with reported study. \cite{sow} Even at high magnetic field, SrRuO$_3$ does not exhibit sign of saturation which is probably due to itinerant nature of magnetism (Fig. 6a). At 70 kOe, we obtain magnetic moment $\mu_H$ = 1.25 $\mu_B$/f.u. which is less than the calculated value (gS$\mu_B$) = 2 $\mu_B$/f.u. with total spin $S$ = 1 and g = 2 is the Lande g-factor. \cite{cao} The variation of remnant magnetization ($M_r$) has given in Table 2. The inset of Fig. 6a shows $H_c$ initially decreases up to $x$ $\sim$ 0.1, and after that it increases. This behavior of $H_c$ appears to be connected with competition between site dilution caused by nonmagnetic Ga$^{3+}$ and mixed ionic states of Ru$^{4+}$ and Ru$^{5+}$. The inset of Fig. 6b shows experimentally estimated $\mu_H$ at 70 kOe with Ga substitution. The observed values of $\mu_H$ decrease linearly with Ga doping in SrRu$_{1-x}$Ga$_x$O$_3$ series, which are expected from this site dilution.

The nature of magnetic state in SrRuO$_3$ and its evolution with Ga substitution has been checked with Arrott plot which is about plotting of the magnetic isotherm $M (H)$ data in the form of $M^2$ vs $H/M$. \cite{arrott} The Arrott plot has been constructed using $M (H)$ data from Fig. 6a. The significance of the Arrott plot is that positive intercept on $M^2$ axis due to straight line fitting in high field regime gives spontaneous magnetization ($M_s$). Arrott plot in Fig. 6b shows non-linear behavior with concave nature. However, the straight lines fitted in high field regime give positive intercept for all the samples. From intercept we calculate $M_s$ = 0.846 $\mu_B$/f.u. for SrRuO$_3$. With Ga doping, $M_s$ decreases which is again in agreement with dilution of magnetic lattice and has shown in the inset of Fig. 6b. 

\begin{figure}
	\centering
		\includegraphics[width=8.5cm]{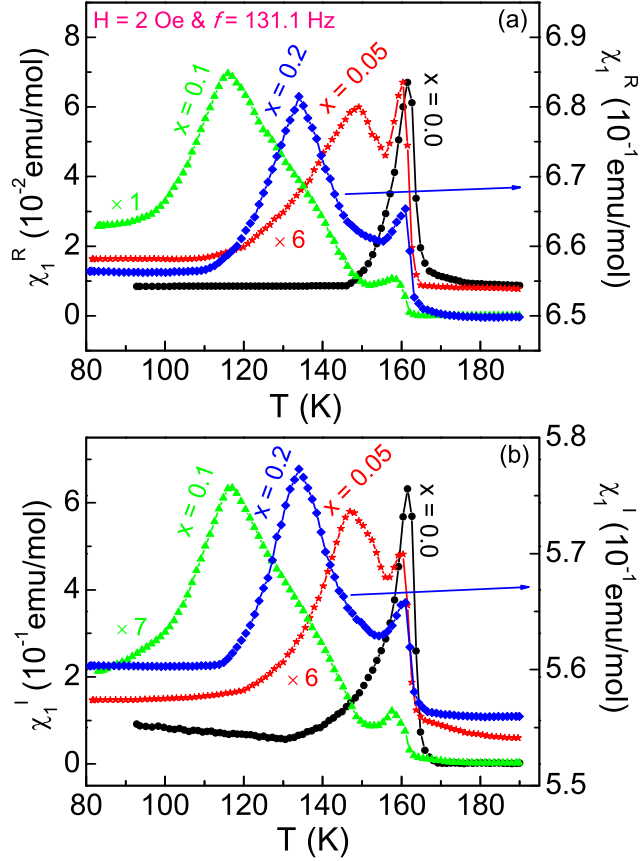}
	\caption{(a) The real ($\chi_1^R$) and (b) imaginary ($\chi_1^I$)	part of ac susceptibility measured in ac field 2 Oe and frequency 131.1 Hz are shown as a function of temperature for SrRu$_{1-x}$Ga$_x$O$_3$ series with $x$ = 0.0, 0.05 and 0.1 (left axis) and $x$ = 0.2 (right axis).}
	\label{fig:Fig7}
\end{figure}

Further, the Rhodes-Wohlfarth (RW) ratio \cite{rhodes} has been used to understand the itinerant character of magnetism in present series. This ratio is to calculate $q_c$/$q_s$ while $q_c$ characterizes the number of carriers per magnetic atom and is derived from the PM state, $\mu_{eff}$ = $g\sqrt{S(S+1)}$ in which $S$ = $q_c/2$, and the $q_s$ represents $\mu_H$ which is the saturated ferromagnetic moment per magnetic atom. It is argued that for $q_c$/$q_s$ $>$ 1, the nature of magnetic interaction would be of itinerant type and the $q_c$/$q_s$ $\sim$ 1 implies localized nature of magnetic state in a material. For SrRuO$_3$, we obtain $q_c$ = 1.82 and $q_s$ = 1.25, therefore the $q_c$/$q_s$ ratio (1.46) favors an itinerant magnetic nature. Table 2 shows the $q_c$/$q_s$ ratio increases with Ga doping which implies that the itinerant ferromagnetic character increases with progressive doping of Ga. 

\subsection{AC susceptibility measurements}
In addition to a prominent peak around $T_c$, the $M_{ZFC}$ in the inset of Fig. 2a shows an extra peak at low temperature in doped samples. Considering the fact that present site dilution in SrRu$_{1-x}$Ga$_x$O$_3$ generates magnetic clusters (GP behavior in Fig. 5), these peaks are likely due to freezing of these clusters. To probe the low temperature magnetic state in present series, we have used ac susceptibility ($\chi_{ac}$) which is an effective tool to understand the nature of magnetic state. \cite{sinha, chakravarti, bitoh, suzuki} Unlike dc magnetization study, the advantage of $\chi_{ac}$ is that it uses very low magnetic field. Moreover, the dynamic nature of magnetic state can be probed through a variation of measurement frequency ($f$) which allows to use different measurement probe time $\tau$ (= $f^{-1}$).

We have measured the $\chi_{ac}$ for all the samples in temperature range covering magnetic phase transition (80 to 200 K). Figs. 7a and 7b show the real ($\chi_1^R$) and imaginary ($\chi_1^I$) part of 1st harmonic $\chi_{ac}$, respectively measured in an ac field $H_{ac}$ = 2 Oe and $f$ = 131.1 Hz. Both $\chi_1^R$ and $\chi_1^I$ show a distinct sharp peak around $T_c$ of respective samples. The $\chi_1^R$ and $\chi_1^I$ further show an additional peak for doped materials below $T_c$ where the peak temperatures closely match with those seen in dc magnetization (Fig. 2a). It is remarkable that the temperatures where an additional peak is observed below $T_c$, substantially agree in case of ac and dc magnetization. 

To further understand the origin of low temperature magnetic peak, we have measured both $\chi_1^R$ and $\chi_1^I$ with varying frequency which allows to probe the system using different time scale. In case of metamagnetic systems such as, spin glass (SG) or super-paramagnets (SPM), the peak temperature ($T_P$) shifts to higher temperature with increasing frequency. \cite{lin-sg, bakaimi, khurshid, anand, yu} Fig. 8a shows an expanded view of $\chi_1^I$ around the low temperature peak for $x$ = 0.1 sample where the data are shown for selected few frequencies. As evident in figure, with increasing frequency the $\chi_1^I$ decreases and the $T_P$ shifts toward higher temperature. While this behavior of $\chi_{ac}$ is typical of glassy or SPM systems but the shift of $T_P$ in present data is very small. \cite{lin-sg, bakaimi} For instance, $T_P$ shifts from 116.06 K to 117.37 K with an increase of frequency from 131.1 Hz to 931.1 Hz. Usually, the frequency response of $T_P$ is quantified with following relation, \cite{lin-sg, anand, mydosh, hohenberg}

\begin{eqnarray}
   \Phi_f = \frac {\Delta T_P} {T_P \Delta {\log_{10}(f)}}
\end{eqnarray}

where $\Delta$ is the difference in respective parameters. In case of glassy and interacting particle systems, the shifting of $T_P$ is very low and the values of $\Phi_f$ is usually obtained in the range of the 0.005 to 0.05. For instance, obtained values of $\Phi_f$ are around 0.005 for canonical spin-glass systems (CuMn, AuMn), \cite{anand, mydosh, mulder, xu, maji} 0.008 for short-range FM system U$_2$RhSi$_3$, \cite{li} 0.002 for interacting manganite nanoparticles, etc. \cite{pramanik} On the other hand, a larger value of $\Phi_f$ ($\sim$ 0.1) is observed for simple SPM systems (i.e., $\Phi_f$ $\sim$ 0.28 for Ho$_2$O$_3$). \cite{mydosh} This suggests an interaction among the magnetic entities gives a low value of $\Phi_f$. For cluster glass systems, where the locally ordered FM clusters interact and play similar role of single spin as in SG systems, $\Phi_f$ is obtained in an intermediate range. \cite{chakrabarty} For present SrRu$_{0.9}$Ga$_{0.1}$O$_3$, we obtain $\Phi_f$ = 0.013 which is significantly low and close to the value for cluster glass systems. This implies a reasonable interaction among the clusters which induces glassy behavior in the system.     

\begin{figure}
	\centering
		\includegraphics[width=8.5cm]{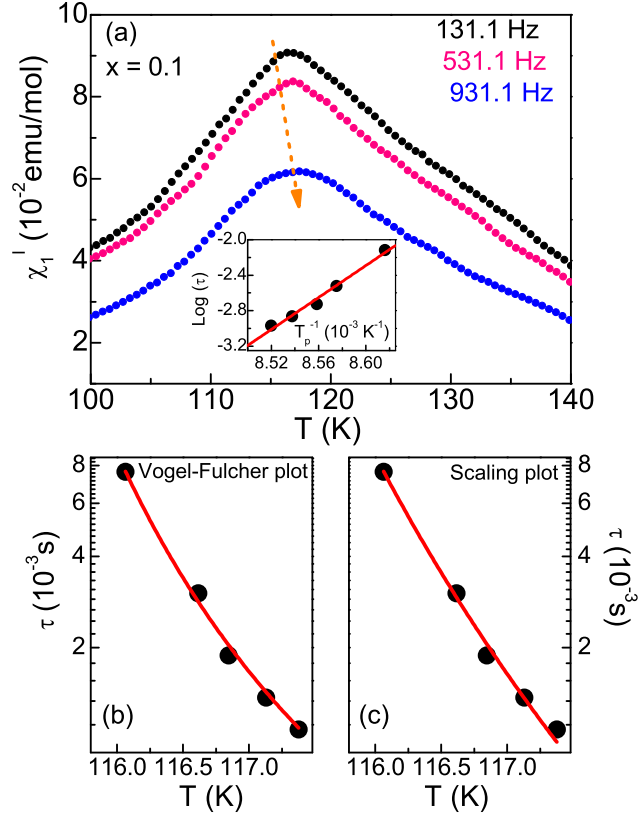}
	\caption{(a) The imaginary part of ac susceptibility ($\chi_1^I$) has been shown as a function of temperature for SrRu$_{0.9}$Ga$_{0.1}$O$_3$ at different frequencies. Inset shows the  N$\acute{e}$el Arrhennius fitting (Eq. 11) of frequency dependent peak temperature $T_P$ (b) Fitting due to Vogel-Fulcher law Eq. 12 and (c) power- law fitting Eq. 13 are shown.}
	\label{fig:Fig8}
\end{figure}

To understand the nature of inter-cluster interactions, we have analyzed the frequency dependent shifting of $T_P$ using different phenomenological models. For an assembly of noninteracting particles/clusters, N$\acute{e}$el Arrhennius law predicts relaxation time ($\tau$) as following, \cite{mydosh, binder-f}

\begin{eqnarray}
   \tau = \tau_0 exp\left(\frac{E_a}{k_B T}\right)
\end{eqnarray}

where ${\tau_0}$ is the microscopic relaxation time, $k_B$ is the Boltzmann constant and $E_a$ is the average thermal activation energy. The inset of Fig. 8a shows the straight line fitting of $\tau$ vs $T_P$ data following Eq. 11. We obtain the fitting parameters $\tau_0$ and $E_a/k_B$ around 2.1 $\times$ 10$^{-35}$ Hz$^{-1}$ and 9017.7 K, respectively which appear to be quite unphysical though the fitting looks very good. \cite{pramanik, peddis} For simple SPM relaxation, the $\tau_0$ comes in the range of 10$^{-8}$ to 10$^{-13}$ s. \cite{pramanik} This implies a presence of sizable interaction among the FM clusters. \cite{gatteschi} We have further tried to analyze the frequency dependance of $T_P$ using Vogel-Fulcher (VF) law, \cite{mydosh, vogel, fulcher, tholence, souletie, shtrikman}

\begin{eqnarray}
   \tau = \tau_0 exp\left[\frac{E_a}{{k_B} (T-T_0)}\right]
\end{eqnarray}

This VF law is an improvement of previous N$\acute{e}$el Arrhennius law where the parameter $T_0$ (0 $<$ $T_0$ $<$ $T_f$) characterizes the inter-cluster interactions. The VF fitting (Eq. 12) of experimental $\tau$ vs $T_P$ data for $x$ = 0.1 sample is shown in Fig. 8b. The fitting yields $\tau_0$ = 1 $\times$ 10$^{-5}$ s, $E_a$/$k_B$ = 20.4 K and $T_0$ = 112.9 K. The high value of $T_0$ implies the strength of interaction is quite significant. Further, $\tau_0$ appears to be very high. However, similar values of $\tau_0$ has been observed in cluster-glass like systems such as, Ni$_2$Mn$_{1.36}$Sn$_{0.64}$ and Ga$_2$MnCo suggesting the slow spin dynamics in present material is due to inter-cluster interactions. \cite{anand, chatterjee, samanta}

The dynamical spin nature has been further investigated using conventional scaling law analysis which assumes that the relaxation time $\tau$ is related to spin coherence length $\xi$. As one approaches the glass transition temperature $T_g$, the $\xi$ diverges suggesting following power-law behavior, \cite{mydosh, hohenberg, binder-f, souletie}

\begin{eqnarray}
   \tau = \tau_0 ({T/T_g}-1)^{-z\nu}
\end{eqnarray}

where $\tau_0$ is the microscopic relaxation time of fluctuating spins or clusters, $z$ is the dynamic scaling exponent and $\nu$ is usual critical exponent related to $\xi$. The solid line in Fig. 8c represents the best fitting of our experimental data for $x$ = 0.1 using Eq. 13. The obtained fitting parameters are $\tau_0$ = 7 $\times$ 10$^{-12}$ s, $z\nu$ = 5.76 and $T_g$ = 113 K. However, the values of $\tau_0$ obtained in VF and scaling analysis, show large disagreement. Nonetheless, $\tau_0$ obtained in scaling analysis lies in the range of typical spin-glass systems i.e., 10$^{-11}$ - 10$^{-12}$ s which suggests that the slow spin dynamics is due to presence of interacting clusters, rather than individual clusters. \cite{lin-sg, bakaimi, anand, yu} Further, obtained exponent $z\nu$ is close to the value for canonical spin-glass systems (i.e., $z\nu$ = 5 - 10). \cite{souletie, fischer} It can be mentioned that the value of $z\nu$ has been theoretically predicted to be 4 for 3D Ising glass system, but in experimental situation $z\nu$ shows a wide variation for different kinds of materials. \cite{binder-t} The fitting as well as obtained values due to scaling law hypotheses (Eq. 13) imply a collective inter-cluster interaction which induces critically slow down spin dynamics. The VF and scaling analysis confirm a cluster-glass like behavior in doped sample which is in agreement with GP behavior.

\begin{figure}
	\centering
		\includegraphics[width=8.5cm]{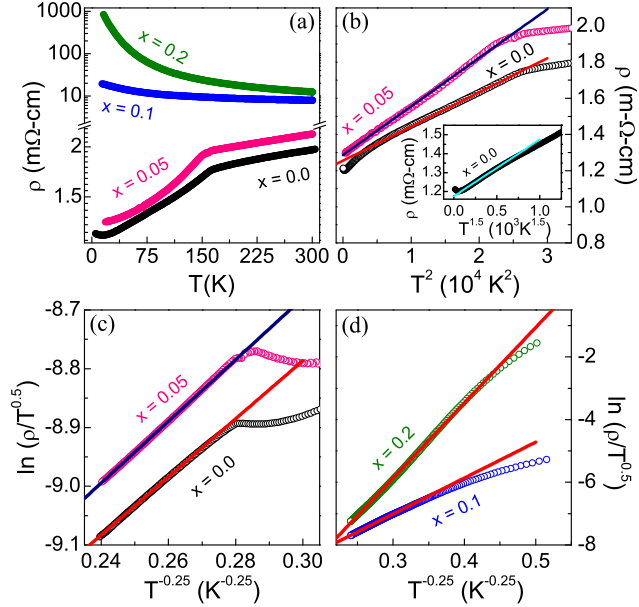}
	\caption{(a) The temperature dependent electrical resistivity have been shown for SrRu$_{1-x}$Ga$_x$O$_3$ series (b) Resistivity in form of $\rho$ vs. $T^2$ are plotted following FL behavior for $x$ = 0 and 0.05 samples. Solid lines are due to straight line fitting. Inset shows $\rho$ verses $T^{1.5}$ showing NFL behavior for $x$ = 0 sample (c) and (d) show the plotting of $\rho (T)$ data following Eq. 15 for $x$ = 0.0, 0.05 and $x$ = 0.1, 0.2 samples respectively. Straight line fittings confirm the modified Mott's VRH model for charge transport}.
	\label{fig:Fig9}
\end{figure}

\begin{figure}
	\centering
		\includegraphics[width=8.5cm]{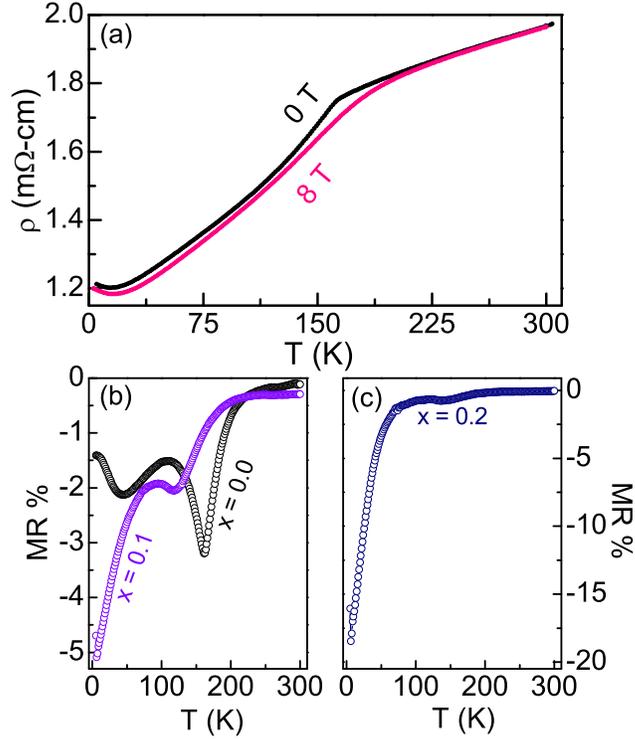}
	\caption{(a) Temperature dependent electrical resistivity $\rho (T)$ measured in 0 and 8 T of magnetic field are shown for SrRuO$_3$. (b) and (c) show the MR\% as function of temperature range from 5 K to 300 K for $x$ = 0.0, 0.1 and $x$ = 0.2 respectively of SrRu$_{1-x}$Ga$_x$O$_3$ series.}
	\label{fig:Fig10}
\end{figure}

\subsection{Electrical resistivity of SrRu$_{1-x}$Ga$_x$O$_3$ series}
The electrical resistivity as a function of temperature $\rho (T)$ have been shown in Fig. 9a for SrRu$_{1-x}$Ga$_x$O$_3$ series. The semi-log plotting of $\rho (T)$ for SrRuO$_3$ shows a metallic behavior with very low resistance value in whole temperature range.\cite{allen} The $\rho (T)$ for SrRuO$_3$, shows a slope change around its $T_c$ where below $T_c$, $\rho (T)$ decreases with faster rate which is probably due to reduced spin disorder effect in magnetically ordered state. The Ga doping in present series increases resistivity and induces an insulating state for $x$ = 0.1 and above, as seen in Fig. 9a. 

Here, it can be mentioned that similar metal to insulator transition (MIT) has been observed SrRu$_{1-z}$Ti$_z$O$_3$ series around $z$ $\sim$ 0.4. \cite{kim-mit} This shows Ga$^{3+}$ has comparatively stronger effect on the electron transport behavior in SrRuO$_3$ inducing insulating behavior at lower concentration of doping. Given that Ga$^{3+}$ and Ti$^{4+}$ causes similar structural modification and both are 3$d$ based nonmagnetic elements. But, Ga$^{3+}$ has fully filled d-orbitals, and it additionally creates Ru$^{5+}$ ion. The Ru$^{4+}$ and Ru$^{5+}$ has different electron configuration (4$d^{4}$ and 4$d^{3}$) which will populate $t_{2g}$ orbitals differently. Therefore, whether this stronger effect of Ga$^{3+}$ on electronic properties is related to mixed charge states of Ru needs to be investigated.

Considering the metallic charge transport for $x$ = 0.0 and 0.05 materials (Fig. 9a), we have examined the charge transport mechanism by plotting the resistivity as a function of $T^2$ following conventional Fermi liquid (FL) theory. For SrRuO$_3$, Fig. 9b shows a linear behavior in $\rho$ vs $T^2$ in temperature range from $T_c$ down to $\sim$ 67 K, which confirms a FL behavior in SrRuO$_3$. It is, however, interesting to observe in Fig. 9b that in Ga doped sample (5\%), the FL behavior is obeyed in whole temperature range below $T_c$. Following the deviation from FL behavior at low temperature in SrRuO$_3$, we have further plotted $\rho$ as a function of $T^{1.5}$ in inset of Fig. 9b. The linear behavior in figure indicates a non Fermi liquid (NFL) type charge conduction prevails at low temperature between around 70 to 20 K. In contrast, previous studies mostly have shown FL like charge transport at low temperature in SrRuO$_3$ which changes to NFL behavior at high temperature that extends up to $T_c$. \cite{klein1, wang, klein-b, tyler, hussey, bruin, schneider} This break down of FL behavior at high temperature has mostly been ascribed to the bad metallic character of SrRuO$_3$ which realized as the $\rho (T)$ shows a linear increase above $T_c$ without any saturation, crossing the Ioffe-Regel limit for conductivity. \cite{klein-b, ziese, allen, gurvitch, klein1, emery} The $\rho (T)$ increases continuously till temperature as high as $\sim$ 1000 K. \cite{allen} In that scenario, the breakdown of FL behavior at low temperature in present study in quite intriguing. This material neither known to be close to any quantum critical point nor it has antiferromagnetic type fluctuations, as in CaRuO$_3$. \cite{longo, cao-c} The NFL behavior is observed at low temperature much below $T_c$ where FM moments are usually firmly ordered due to reduced thermal fluctuations. We understand that this NFL behavior arises due to scattering of itinerant electrons with magnetic moments. In doped sample, where the magnetism is weakened due to substitution of Ga, the FL behavior continues to low temperature. Nonetheless, continuation of FL behavior down low temperature with only 5\% of Ga is intriguing.
 
The nature of $\rho (T)$ in PM state is different for $x$ = 0.1 and 0.2 (Fig. 9a) samples in present series. Generally, charge transport mechanism in an insulating state follows Mott’s variable range hopping (VRH) model which is expressed in case of 3-dimensional material as, \cite{mott}   

\begin{eqnarray}
	\rho (T) = \rho_0  exp[(T_M /T)^{1/4}]
\end{eqnarray}
  
where $\rho_0$ and $T_M$ are the constants. This formula would then be valid for higher doped insulating samples of SrRu$_{1-x}$Ga$_x$O$_3$ series. However, the present samples have reasonable disorder arising from nonmagnetic Ga$^{3+}$ introduced at active Ru-O channel of charge transport. The Eq. 14 has been further modified by Greaves for disordered materials where a $\sqrt{T}$ term has been included as prefactor as following, \cite{greaves}

\begin{eqnarray}
	\rho (T) = \rho_0 \sqrt{T}  exp[(T_M /T)^{1/4}]
\end{eqnarray}

The $\rho_0$ accounts for electron-phonon interaction and has following form       

\begin{eqnarray}
	\rho_0 = \frac{1}{3 e^2 \nu_{ph}} \left[\frac{8 \pi \alpha k_B}{N(E_F)}\right]^{1/2}    
\end{eqnarray}
             
where $\nu_{ph}$ ($\sim$ 10$^{13}$ s$^{-1}$) is an optical phonon frequency, $\alpha$ = 1/$\xi$ is an inverse localization length of the localized states, $k_B$ is the Boltzmann constant, $N$($E_F$) is the density of states and $e$ is the electronic charge. The $T_M$ in Eq. 15 is the measure of disorder present in the system and can be written as			

\begin{eqnarray}
	T_M = \Lambda \left[\frac{\alpha^3} {k_B N(E_F)}\right]  
\end{eqnarray}
			
where $\Lambda$ is constant and has value 19.45 for disordered state. \cite{sakata, dutta} The $\rho (T)$ data are fitted with Eq. 15 for all the samples in high temperature PM range. Figs. 9c and 9d show the fitting of $\rho (T)$ data following Eq. 15 for $x$ = 0.0, 0.05 and $x$ = 0.1, 0.2 samples, respectively. It is worth to mention here that charge conduction for both metallic and insulating samples can be explained with single model (Eq. 15). The validity of Eq. (15) for charge transport in SrRuO$_3$ has been in early report. \cite{kurnia} The parameters $\rho_0$ and $T_M$ obtained from fitting in Figs. 9c and 9d are given in Table 3. It is clear in Table 3 that while $\rho_0$ does not vary appreciably but $T_M$ shows a large increase with Ga doping which signifies the role of disorder in system arising from site dilution.

\begin{table}
\caption{\label{tab:table III} Table shows the values of coefficients $\rho_0$ and $T_M$ related to modified Mott's VRH model (Eq. 15) and the exponents $p$ and $q$ for a magnetoresistance (Eq. 20 and 21) for SrRu$_{1-x}$Ga$_x$O$_3$ series.}

\begin{indented}
\item[]\begin{tabular}{cccccc}
\hline
Quantity/Ga(x) &0.0 &0.05 &0.1 &0.2\\
\hline
$\rho_0$ (m$\Omega$-cm $K^{-1/2}$) &0.037 &0.035 &0.031 &0.003\\
\hline
$T_M(K)$ &466 &625 &15345 &243478\\
\hline
$p$ &0.9 & &0.95 &1.56\\
\hline
$q$ &0.3 & &0.28 &0.4\\
\hline
$p/q$ (Eq. 20) &3 & &3.44 &4\\
\hline
$p/q$ (Eq. 21) &3.34 & &4 &5.8\\
\hline
\end{tabular}
\end{indented}
\end{table}

\begin{figure}
	\centering
		\includegraphics[width=8.5cm]{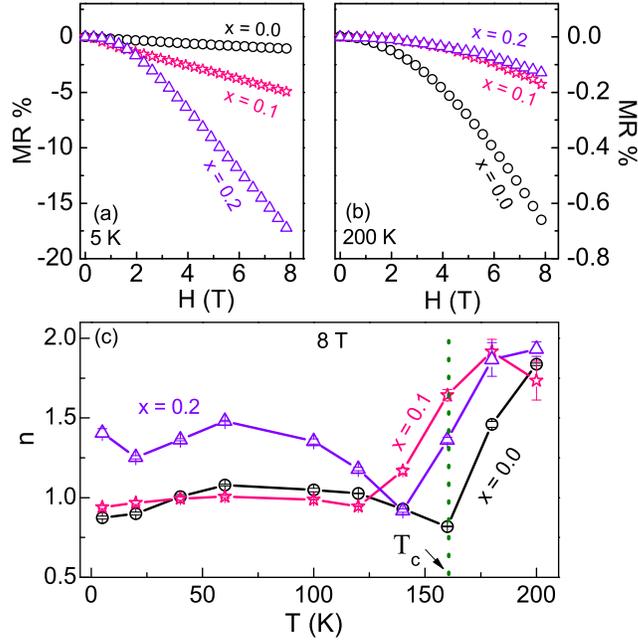}
	\caption{Magnetic field dependent isothermal MR are shown for compound of SrRu$_{1-x}$Ga$_x$O$_3$ series at (a) 5 K (b) 200 K (c) The exponent $n$, due to fitting of MR data with Eq. 19 is shown with temperature for $x$ = 0.0, 0.1 and 0.2 materials of SrRu$_{1-x}$Ga$_x$O$_3$ series. The vertical dotted line represent $T_c$ of SrRuO$_3$.}
	\label{fig:Fig11}
\end{figure}

\subsection{Magnetoresistance Study}
The temperature dependent resistivity in presence of magnetic field (80 kOe) has been shown in Fig. 10a for SrRuO$_3$. It is observed in figure that in presence of field the overall feature of $\rho (T)$ does not change, however, the value of $\rho$(T) decreases in FM state. This indicates that a negative magnetoresistance (MR) in SrRuO$_3$, as observed in other studies. \cite{sow, moran} The prominent effect of MR is observed across $T_c$ as the FM transition is broadened in high magnetic field. The magnitude of MR is usually defined as the percentage change in $\rho$ in the presence of magnetic field as,  

\begin{eqnarray}
	MR \% = \left[\frac {\rho (H) - \rho (0)} {\rho (0)}\right] \times 100
\end{eqnarray}  

where $\rho (H)$ and $\rho (0)$ are the resistivity measured in magnetic field and zero field, respectively. The temperature dependent MR have shown in Figs. 10b and 10c for $x$ = 0.0, 0.1 and $x$ = 0.2 respectively indicates a negative MR for all the materials over the temperature range. For SrRuO$_3$ MR is not substantial but it shows dip around $T_c$ and at low temperature around 50 K that also matches with other study. \cite{sow} This is interesting as we find electronic transport deviates from Fermi liquid behavior below around 70 K (Fig. 9b). The negative MR increases substantially for insulating samples $x$ = 0.1 and 0.2 material reaches around 5\% and 18\% respectively at 5 K (Fig. 10c) of SrRu$_{1-x}$Ga$_x$O$_3$ series. This substantial increase of MR at low temperature is probably due to a high insulating state. This material continue to be ferromagnetic and the alignment of spins in field promotes the charge conduction. It can be further noticed in Fig. 10 that MR for both $x$ = 0.1 and 0.2 shows a dip around 118 K and 135 K respectively which are the glass transition temperature for both these samples.

The isothermal MR data at 5 K and 200 K are further plotted in Fig. 11a and 11b respectively where both the temperature represents magnetic and nonmagnetic state, respectively. As discussed above, negative MR at 5 K increases substantially from $\sim$ 1\% to $\sim$ 18\% with increase of Ga from 0 to 20\%. It is further noticeable in Fig. 11a that while SrRuO$_3$ shows almost linear MR with field, a nonlinearity in MR is induced in with increasing Ga doping. This implies that $x$ = 0.2 material is largely diluted, therefore, spin alignment induced conduction happens at higher field. Opposed to low temperature behavior, the MR at 200 K decreases with Ga (Fig. 11b). Further, Fig. 11b shows the field dependent MR at 200 K is nonlinear for all the samples.

The non-linearity in MR has been quantified by fitting with following power-law behavior,

\begin{eqnarray}
	MR = A + B H^n
\end{eqnarray}

where $A$ and $B$ are the constants and $n$ is the power exponent. Fig. 11c shows the exponent $n$ with temperature for $x$ = 0.0, 0.1 and 0.2 samples. Fig. 11c shows that MR for $x$ = 0.0 and 0.1 increases almost linearly ($n$ $\sim$ 1) till $T_c$, then above $T_c$ the value of $n$ increases to $\sim$ 2. The $x$ = 0.2 sample, on the other hand, shows higher power dependance of MR on field even at low temperatures. At 200 K in PM state, the MR shows quadratic field dependance for all the materials. 

\begin{figure}
	\centering
		\includegraphics[width=8.5cm]{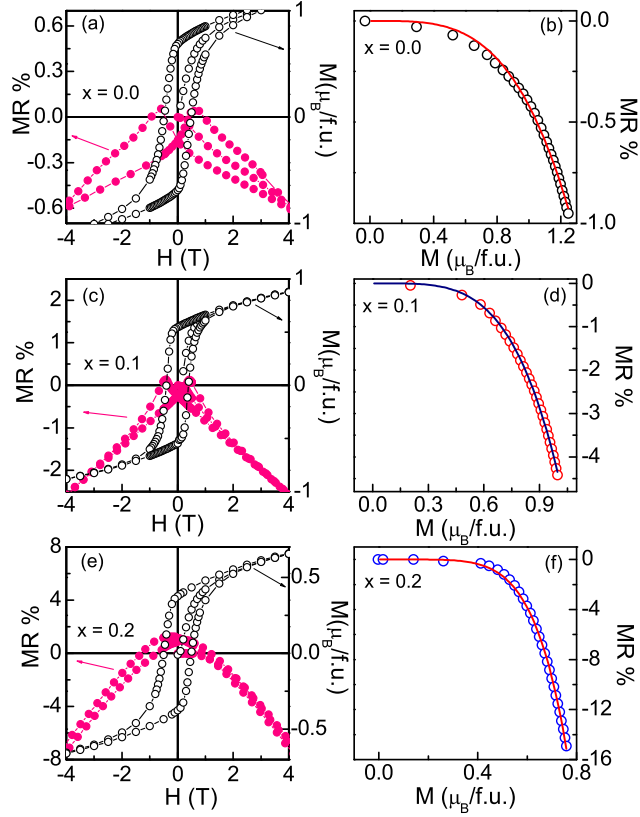}
	\caption{ The magnetoresistance (left axis) and the magnetization (right axis) for (a) $x$ = 0.0 (c) $x$ = 0.1 and (e) $x$ = 0.2 samples are shown at 5 K as a function of magnetic field of SrRu$_{1-x}$Ga$_x$O$_3$ series. (b), (d) and (f) show the MR and corresponding moment at same field which have been deduced from (a), (b) and (c) for $x$ = 0.0, 0.1 and 0.2 samples, respectively. The solid lines are due to fitting with Eq. 21.}
	\label{fig:Fig12}
\end{figure}

Considering that magnetic ordering has sizable influence on MR behavior, we have plotted the field dependent MR along with $M (H)$ data at 5 K. Figs. 12a, 12c and 12e show the combined MR and magnetization data at 5 K for $x$ = 0.0, 0.1 and 0.2, respectively. While all the materials  exhibit a negative MR, but its nature vary with the effect of site dilution. 

For example, MR for parent SrRuO$_3$ increases during first application of magnetic field but when the field is returned back to zero, the MR shows significant hysteresis and remnant value. When the field is switched to negative direction, MR continues to decrease and shows a positive peak at field which roughly coincides with its coercive field (Fig. 12a). With further increase of negative field, the MR increases toward negative direction and exhibits a hysteresis with reversal of field. Important observations are MR exhibits significant hysteresis, a remnant MR at zero field and the positive peaks around coercive fields, where all appear to be connected with the magnetic state of material. The spin alignment with increasing field promotes the charge conduction which results in negative MR. The remnant negative MR is caused by remnant magnetization at zero field which is evident in $M (H)$ plot in Fig. 12a. When the field is increased to negative direction, the negative MR decreases and MR shows zero value at field that closely matchs with the $H_c$. 

The slight positive MR immediately after $H_c$ is likely due to effect of field as the moment is close to zero. However, spin alignment with negative field again induces negative MR which explains the hysteresis in MR. With Ga substitution, we observe similar MR behavior for $x$ = 0.1 but the hysteresis in MR is reduced. 

For $x$ = 0.2, hysteresis in MR is reduced and remnant in MR decreases substantially. Instead, MR shows positive value for larger field range close to zero field. This reduced hysteresis and increased positive MR is likely due to magnetic dilution with nonmagnetic Ga.

We have further attempted to establish a direct correlation between MR and magnetization data. With following functional dependence of MR and $M$ on field $H$, \cite{nath}         

\begin{eqnarray}
   MR \propto H^p, \ \ \     M \propto H^q
\end{eqnarray} 

where $p$ and $q$ are the exponents, a direct relationship between MR and M as be obtained as shown below, 

\begin{eqnarray}
   |MR| \propto M^{p/q}
\end{eqnarray}
 
In Figs. 12b, 12d and 12f, we have plotted MR vs $M$ at 5 K for $x$ = 0.0, 0.1 and 0.2, respectively. The data have been fitted with Eq. 21 which yields $p/q$ = 3.34 for SrRuO$_3$. Using Eq. 20, we separately obtain $p$ = 0.9 and $q$ = 0.3 which implies a $p/q$ ratio = 3 for SrRuO$_3$. The both $p/q$ values obtained from Eqs. 20 and 21 for this series (shown in Table 3) indicate a close matching. Here, it can be mentioned that the negative MR has been shown to scale as $\propto$ (${M/M_s}$)$^3$ for itinerant FM system Mn$_x$TiS$_2$ while for localized moment systems the dependance is quadratic.\cite{negishi} The present $p/q$ exponent ($\sim$ 3) for SrRuO$_3$ shows a nice agreement with itinerant system. Further, an increase of $p/q$ exponent with $x$ (Table 3) implies an increasing trend of itinerant character in SrRu$_{1-x}$Ga$_x$O$_3$ which also supports the results, as obtained from increasing Rhodes-Wohlfarth ratio (Table 2). 

\section{Conclusion}
Series of polycrystalline samples SrRu$_{1-x}$Ga$_x$O$_3$ with $x$ = 0.0, 0.05, 0.1 and 0.2 are prepared using solid state method to understand the complex magnetic and transport properties in SrRuO$_3$. All the samples are characterized with orthorhombic-\textit{Pbnm} crystal structure. However, the lattice parameters ($a$, $b$ and $c$) progressively decrease with $x$ which is due to formation of Ru$^{5+}$ that has smaller ionic radii compared to Ru$^{4+}$ and Ga$^{3+}$. Both the magnetic moment as well as FM $T_c$ decreases with Ga$^{3+}$ substitution. The analysis of thermal demagnetization at low temperature indicates a coexistence of both itinerant and localized model of magnetization. The estimated magnetic critical exponent $\beta$ indicates a mean-field type magnetization in SrRuO$_3$. The value of $\beta$ increases with Ga that is ascribed to site dilution effect. This site dilution leads to formation of FM clusters which is evident through Griffiths phase like behavior in higher doped samples. The ac susceptibility measurements confirm that a cluster-glass like behavior at low temperature that is also induced by FM cluster formation. A transition from metallic to insulating state is observed between $x$ = 0.05 to 0.1. The charge transport behavior in both high temperature PM region of $x$ = 0 and 0.05 and in insulating states of $x$ = 0.1 and 0.2 is explained by modified Mott's VRH model. But in FM-metallic state just below $T_c$ ($x$ = 0 and 0.05), the charge transport mechanism mainly follows Fermi-Liquid behavior. At low temperature, a breakdown in FL behavior occurs due to an increased scattering of itinerant electrons by ordered moment. The site dilution effect is also evident in charge transport as doped sample shows FL behavior throughout the FM state. 

\section{Acknowledgment}   
We acknowledge UGC-DAE CSR, Indore, Dr. Alok Banerjee and Dr. Rajeev Rawat for the for magnetization and transport measurements. We thank Mr. Kranti Kumar and Mr. Sachin for the help in measurements. We are also thankful to AIRF, JNU and Mr. Saroj Jha for magnetization measurements. We thank DST-PURSE for the financial supports. RG and INB acknowledge the financial support from UGC, India (BSR fellowship) and from CSIR, India (SRF Fellowship), respectively.

\end{document}